# FAST ADAPTIVE NEURAL CONTROL OF RESONANT EXTRACTION AT FERMILAB

A. Whitbeck*, J. Berlioz, K. Danison-Fieldhouse, K. Hazelwood, M. Khan,
J. Mitrevski, A. Narayanan, J. St. John, N. Tran, Fermilab, Batavia, IL, USA
J. Ji, M. Walter, Toyota Technical Institute of Chicago, Chicago, IL, USA

*Abstract*

We present progress on the development of a machine learning (ML) regulation system for third-order resonant extraction of the beam delivered to the Mu2e experiment at Fermilab. We consider classical and ML-based controllers optimized on semi-analytic simulations and provide performance comparisons for several models. Additionally, we discuss the efficiency of each model in training, which has implications for future work on adaptive control. We also discuss progress on developing optimized implementations of ML models for edge-based inference.

## INTRODUCTION

The Fermilab Muon Delivery Ring provides proton beam over many turns to the muon production target for the Mu2e experiment using a third-order resonant extraction. The tune of the machine is controlled using a series of fast-ramping quadrupole magnets. While the nominal quadrupole ramp waveform is sourced from a spill-by-spill optimization loop, the fast disturbances in the system are regulated by a dedicated hardware-based controller. The controller reads the digitized signal from the wall current monitor in the M4 extraction beamline to track the extracted beam current and provides updates to the quad controller at 10 kHz. Controllers will be run in a field-programmable gate array (FPGA) to meet the latency requirements of the system. Here, we extend previous work [1] and explore possible controller implementations, evaluate their performance, and compare optimization strategies. The requirements of the system and further details of the Mu2e Spill Regulation System (SRS) are described in Ibrahim *et al.* [2, 3].

## DATA AND METHODOLOGIES

Our study builds on an analytically derived model that efficiently approximates spill dynamics [1, 3]. Spill quality is quantified by the *spill-duty factor* (SDF), SDF = $(1 + \sigma_{\text{spill}}^2)^{-1}$, where $\sigma_{\text{spill}}$ is the standard deviation of the normalised spill intensity $\{s_t\}_{t=1}^T$. The goal of the regulator is to maximize this objective.

Each episode represents a single spill of fixed horizon $T = 430$ control cycles. At step $t$, the agent observes the one-dimensional state $s_t$ and outputs an action $a_t \in \mathbb{R}$, which is added to the pre-programmed quadrupole-current ramp. State transitions are generated by the analytical spill model. Figure 1 illustrates the spill dynamics, comparing raw and corrected spill intensities along with the corresponding control signals.

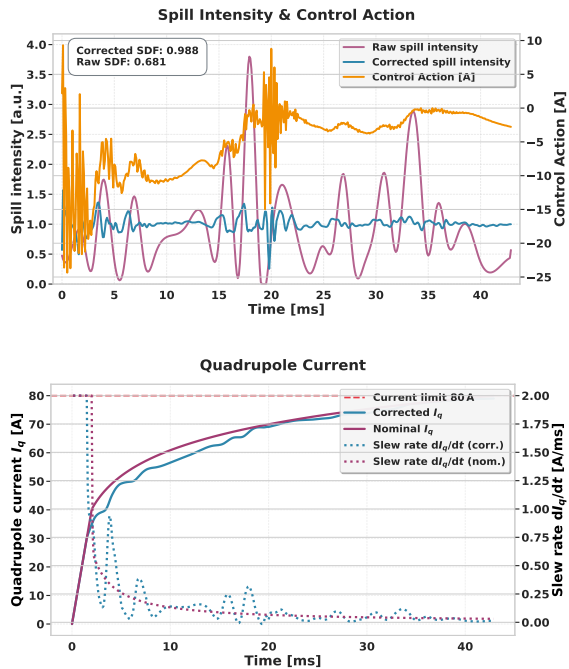

Figure 1: (Upper) Time history of the raw spill intensity versus the corrected spill intensity after feedback during a single 43 ms extraction. The overlaid control signal (right axis) is the current increment commanded to the quadrupole magnet every 0.1 ms. (Bottom) The quadrupole current and its rate of change. The magenta line tracks the nominal ramp $I_q$, the blue line represents the corrected $I_q$ respecting the hardware limit of 80 A (red dashed), and the dotted lines give the corresponding slew rate $dI_q/dt$ per 0.1 ms step.

We have re-implemented the existing differentiable PyTorch simulator to support fully-batched operations [1, 4], enabling the concurrent simulation of thousands of spills on modern GPUs and significantly reducing training time. In addition, we provide an Farama Gymnasium-compatible wrapper [5] that standardises the interface between the agent and the simulator, thereby streamlining the development and comparative evaluation of control algorithms, especially those based on reinforcement learning.

In this work, we investigate two primary classes of controllers: the classical proportional-integral-derivative (PID) model and models based on neural networks (NN). Each

---

provides a distinct trade-off between interpretability and performance.

**PID Controller** The baseline controller is a PID algorithm. Its primary advantage is its high degree of interpretability—the proportional ($G_p$), integral ($G_i$), and derivative ($G_d$) gains have clear physical meanings, making the controller's behavior easier to understand. However, its performance can be limited when controlling complex, nonlinear systems. To establish a robust baseline, the PID gains were optimized using two distinct methods, both leveraging our batched differentiable simulation environment.

(i) A grid search formulates the PID tuning as a discrete optimization over a discretized and bounded parameter space. To ensure statistical reliability, we evaluate each gain triplet $\mathbf{g}^{(k)}$ across the same set of $M$ different simulated episodes and use the highest average SDF to find the optimal point in the three-dimensional parameter space.

(ii) We also explore gradient descent-based optimization of the PID gains [6], which is more efficient and leverages our fully differentiable simulator to directly find optimal gains. The objective is to find the PID gains $\mathbf{G}$ that minimize a composite loss function, $\mathcal{L}_{\text{total}}$:

$$\mathcal{L}_{\text{total}} = w_1 \mathcal{L}_{\text{spill}} + w_2 \mathcal{L}_{\text{surrogate}} + w_3 \mathcal{L}_{\text{reg}},$$

where $w_i$ are scalar weights for each loss component.

Each term in the loss plays a unique role. The Corrected Spill Loss ($\mathcal{L}_{\text{spill}}$) penalizes the deviation of the corrected spill intensity from an ideal uniform distribution (of value 1.0). It is defined as

$$\mathcal{L}_{\text{spill}} = \frac{1}{N \cdot T} \sum_{i=1}^{N} \sum_{t=1}^{T} \tanh\left(\frac{|I_{\text{corr}}^{(i)}(t) - I_{\text{ref}}|}{C}\right).$$

Here, $N$ is the number of spills in a batch, $T$ is the number of time steps, $I_{\text{corr}}^{(i)}(t)$ is the corrected spill intensity of spill $i$ at time $t$, and $C$ is a scaling constant. The tanh function provides bounded and smooth gradients. The surrogate SDF Loss ($\mathcal{L}_{\text{surrogate}}$) provides a differentiable surrogate for maximizing the SDF by minimizing the spill variance and is defined as the averaged $\log(1 + (\sigma_{\text{spill}}^{(i)})^2)$, where $(\sigma_{\text{spill}}^{(i)})^2$ is the variance of the $i$-th corrected spill in the batch. While directly minimizing variance ($\mathcal{L} = \sigma^2$) is the most direct approach, it is often impractical for training neural networks due to its instability. The surrogate log loss ($\mathcal{L} = \log(1 + \sigma^2)$) achieves the same optimization goal while providing crucial robustness. By "taming" the influence of large errors, it ensures smoother and more reliable gradient flow. Finally, the Regularization Loss ($\mathcal{L}_{\text{reg}}$) stabilizes training by penalizing large changes in the absolute value of the gain changes between steps.

**Neural Network Based Models** To overcome the performance limitations of the PID controller, we explore NN-based models, which are well-known for their powerful representability. These models can handle the intricate, nonlinear relationships in the system that a PID controller cannot.

We train the neural networks using supervised learning. We parameterize the spill controller with a neural network that maps a short history of measured spill intensities to a corrective action. Given a batch of $N$ parallel environments and a trajectory length of $T$ steps, the network produces a sequence of actions applied to the quad magnet, yielding corrected intensities $I_{\text{corr}}^{(i)}(t)$. Training minimizes the mean absolute deviation of each corrected spill from the ideal uniform profile $I_{\text{ref}} = 1$:

$$\mathcal{L}_{\text{SL}} = \frac{1}{N \cdot T} \sum_{i=1}^{N} \sum_{t=1}^{T} |I_{\text{corr}}^{(i)}(t) - I_{\text{ref}}|$$

We evaluate two complementary sequence models: a single-layer Gated Recurrent Unit (GRU) [7, 8] and a one-dimensional Temporal Convolutional Network (TCN) [9]. The former takes one intensity sample per time step, while the latter uses the 60 most recent samples as inputs.

## EXPERIMENTS

A key consideration for long-term control system maintenance is sample efficiency—the data required to adapt to changes in accelerator parameters. To evaluate this, we examined how controller performance scales with training dataset size.

We trained controllers on datasets of 10, 100, and 1000 distinct spills, then evaluated on a fixed set of 1000 unseen spills to assess generalization.

Table 1: Sample-efficiency comparison across training-set sizes. Each entry reports the mean SDF ± standard deviation on the same evaluation set of 1000 unseen spills.

| Method | Number of training spills | | |
| --- | --- | --- | --- |
| | **10** | **100** | **1000** |
| PID+GS | 0.680 ± 0.136 | 0.683 ± 0.137 | 0.683 ± 0.137 |
| PID+GD | 0.636 ± 0.150 | 0.634 ± 0.150 | 0.635 ± 0.150 |
| GRU+SL | 0.719 ± 0.296 | 0.908 ± 0.112 | 0.919 ± 0.092 |
| TCN+SL | 0.646 ± 0.155 | 0.751 ± 0.165 | 0.793 ± 0.145 |

Table 1 presents the sample efficiency results. The PID+GS (grid search) and PID+GD (gradient descent) controllers show no performance gain beyond the initial 10 training spills. In contrast, the supervised learning models exhibit strong sample efficiency. GRU+SL (supervised learning) improves markedly from a mean SDF of 0.719 with 10 spills to 0.919 ± 0.092 with 1000, demonstrating effective scalability. TCN+SL also benefits from more data, though its performance peaks at 500 spills (not shown here) before slightly declining. Overall, GRU+SL best leverages larger datasets for superior control performance.

Figure 2 plots each of the 1000 noisy test spills, with points above the diagonal ($y = x$) indicating successful correction by the controller. Performance varies notably across methods, particularly for spills with severe disturbances (SDF <

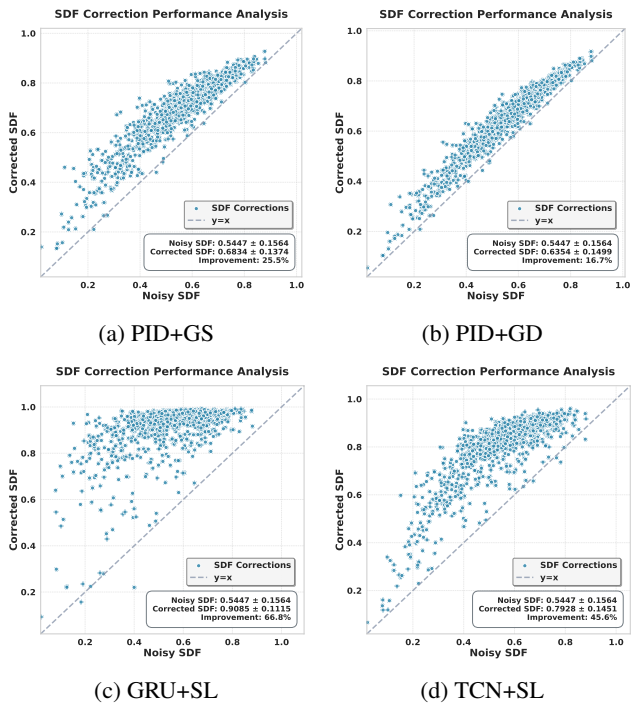

Figure 2: Qualitative comparison of four controllers trained on 500 spills, evaluated on the same set of 1000 noisy spills. Each point plots one spill's SDF before (*x*-axis) vs. after correction (*y*-axis).

Table 2: Influence of action-clipping thresholds on controller performance. Each entry reports the mean SDF ± standard deviation on the same evaluation set of 1000 unseen spills.

| Method | Action-clipping threshold $c_\text{clip}$ (A) | | |
| --- | --- | --- | --- |
| | 3 A | 5 A | 10 A |
| PID+GS | 0.601 ± 0.152 | 0.631 ± 0.151 | 0.649 ± 0.152 |
| PID+GD | 0.570 ± 0.151 | 0.577 ± 0.150 | 0.574 ± 0.150 |
| GRU+SL | 0.768 ± 0.173 | 0.801 ± 0.163 | 0.860 ± 0.143 |
| TCN+SL | 0.763 ± 0.165 | 0.787 ± 0.154 | 0.800 ± 0.136 |

0.4). GRU+SL shows the most significant improvement under these conditions, effectively correcting spills with very low initial SDF values.

To assess the impact of actuator limitations, we trained each model on 100 spills with action-clipping thresholds of ±3 A, ±5 A, and ±10 A, and evaluated on 1000 unseen spills.

Table 2 summarizes controller performance across clipping thresholds $c_\text{clip}$. Learning-based controllers (GRU+SL and TCN+SL) consistently outperform the baseline PID controllers, achieving significantly higher mean SDF scores at all tested thresholds.

Sensitivity to $c_\text{clip}$ varies by method. For GRU+SL, performance improves markedly with larger thresholds, with mean SDF increasing from 0.768 to 0.860 as $c_\text{clip}$ rises from 3 A to 10 A—approaching the performance of the unbounded controller observed in the previous sample-efficiency experi-ment. This suggests that larger action magnitudes are critical for correcting extremely noisy spills.

Deploying the neural controller on the SRS FPGA requires a small memory footprint and sub-millisecond latency [10, 11]. To evaluate how the *number of network parameters* ($|\theta|$) affects control quality, we use GRU+SL as a testbed. Varying the hidden-state dimension (16, 32, 64, 128) yields four model variants—Tiny, Small, Medium, and Large—for performance comparison.

Table 3: Sample-efficiency of GRU+SL models with different numbers of parameters. Each cell shows mean SDF ± standard deviation on a 1000-spill evaluation set.

| Params (K) | Number of training spills | | |
| --- | --- | --- | --- |
| | 10 | 100 | 1000 |
| 2.45 | 0.812 ± 0.141 | 0.823 ± 0.133 | 0.869 ± 0.113 |
| 9.51 | 0.774 ± 0.165 | 0.897 ± 0.116 | 0.902 ± 0.102 |
| 37.44 | 0.719 ± 0.296 | 0.908 ± 0.112 | 0.919 ± 0.092 |
| 148.61 | 0.809 ± 0.139 | 0.912 ± 0.110 | 0.931 ± 0.091 |

Table 3 shows the performance of GRU+SL models of varying sizes on a shared evaluation set, conditioned on different training set sizes. Mean SDF scores increase with more training data across all configurations. For instance, the Large model improves from 0.809 with 10 spills to 0.931 with 1000, while its standard deviation decreases, indicating greater stability.

The benefit of increased model capacity becomes more apparent with larger training sets. Although performance is inconsistent with only 10 spills, models with more parameters (Small, Middle, Large) consistently outperform the Tiny model when trained on 100 or more samples, suggesting that larger models generalize better given sufficient data.

## CONCLUSION

Our work demonstrates the potential of neural network-based controllers for the Mu2e experiment beam simulation. Real-world deployment faces the challenge of reliance on a static, differentiable model. To bridge this sim-to-real gap, we will develop a surrogate of our simulation [12] and later a digital twin that can continuously refine their models of the system dynamics. Ultimately, we aim to leverage the digital twin for efficient, online controller training. By interacting with this fast and accurate digital twin, the neural controller can adapt to system changes in near real-time, enabling robust, self-tuning control systems for real-world deployment.

## ACKNOWLEDGEMENTS

This manuscript has been authored by FermiForward Discovery Group, LLC under Contract No. 89243024CSC000002 with the U.S. Department of Energy, Office of Science, Office of High Energy Physics.